# Guanidinium can both Cause and Prevent the Hydrophobic Collapse of Biomacromolecules

Jan Heyda[1,2,*,+], Halil I. Okur[4, †,+], Jana Hladílková[5,7], Kelvin B. Rembert[4], William Hunn[3], Tinglu Yang[4], Joachim Dzubiella[1,6], Pavel Jungwirth[7]*, and Paul S. Cremer[4,8,]*

[1] Institut für Weiche Materie und Funktionale Materialien, Helmholtz-Zentrum Berlin für Materialien und Energie, Hahn-Meitner Platz 1, 14109 Berlin, Germany

[2] Physical Chemistry Department, University of Chemistry and Technology, Prague, Technicka 5, 16628, Prague 6, Czech Republic

[3] Chemistry Department, Texas A&M University, 3255 TAMU, College Station, Texas 77843, United States.

[4] Chemistry Department, [8] Biochemistry and Molecular Biology Department, The Pennsylvania State University, 104 Chemistry Building, University Park, Pennsylvania 16802, United States

[5] Division of Theoretical Chemistry, Lund University, POB 124, 22 100, Lund, Sweden

[6] Institut für Physik, Humboldt-Universität zu Berlin, Newtonstr. 15, 12489 Berlin, Germany

[7] Institute of Organic Chemistry and Biochemistry, Academy of Sciences of the Czech Republic, Flemingovo nám. 2, 16610 Prague 6, Czech Republic

[+]These authors contributed equally to this manuscript

*Corresponding authors: jan.heyda@vscht.cz (JH), pavel.jungwirth@uochb.cas.cz (PJ), and psc11@psu.edu (PSC).*

**Abstract**

A combination of Fourier transform infrared and phase transition measurements as well as molecular computer simulations, and thermodynamic modeling were performed to probe the mechanisms by which guanidinium ($Gnd^+$) salts influence the stability of the collapsed versus uncollapsed state of an elastin-like polypeptide (ELP), an uncharged thermoresponsive polymer. We found that the cation's action was highly dependent upon the counteranion with which it was paired. Specifically, $Gnd^+$ was depleted from the ELP/water interface and was found to stabilize the collapsed state of the macromolecule when paired with well-hydrated anions such as $SO_4^{2-}$. Stabilization in this case occurred via an excluded volume (or depletion) effect, whereby $SO_4^{2-}$ was strongly partitioned away from the ELP/water interface. Intriguingly, at low salt concentrations, $Gnd^+$ was also found to stabilize the collapsed state of the ELP when paired with $SCN^-$, which is a strong binder for the ELP. In this case, the anion and cation were both found to be enriched in the collapsed state of the polymer. The collapsed state was favored because the $Gnd^+$ crosslinked the polymer chains together. Moreover, the anion helped partition $Gnd^+$ to the polymer surface. At higher salt concentrations (> 1.5 M), GndSCN switched to stabilizing the uncollapsed state because a sufficient amount of $Gnd^+$ and $SCN^-$ partitioned to the polymer surface to prevent cross-linking from occurring. Finally, in a third case, it was found that salts which interacted in an intermediate fashion with the polymer (e.g. GndCl) favored the uncollapsed conformation at all salt concentrations. These results provide a detailed, molecular-level, mechanistic picture of how $Gnd^+$ influences the stability of polypeptides in three distinct physical regimes by varying the anion. It also helps explain the circumstances under which guanidinium salts can act as powerful and versatile protein denaturants.

**Introduction:**

Numerous processes, from ion toxicity to the pickling of cucumbers have been shown to follow the Hofmeister series, a rank ordering of ion specific effects on the physical properties of organic molecules, proteins, and colloids in aqueous solutions containing salts.[1-3] Reversed or inverted Hofmeister series have also been reported.[4-6] More recently, attention has been focused on the underlying molecular level mechanisms involved in these processes in order to shed light on the recurring ion trends.[5,7-9] For example, weakly hydrated and polarizable anions (such as $SCN^-$, $I^-$, or $ClO_4^-$) have been shown to interact with backbone methylene groups in proteins and peptides.[7,8] By contrast, $Li^+$, $Ca^{2+}$, $Mg^{2+}$ and other divalent cations have been shown to interact, albeit very weakly, with the amide oxygen.[9] The specific nature of the anion is usually the predominant factor in determining ion specific Hofmeister effects. Nevertheless, a well-known exception to this observation involves the behavior of $Gnd^+$, which is widely used as a protein denaturant.

Compared to metal cations, $Gnd^+$ has a unique molecular structure (Figure 1). It possesses flat hydrophobic faces, yet it is capable of directional H-bonding along its edges via three $NH_2$ groups. Because of this dual character, $Gnd^+$'s physico-chemical behavior is very unusual. For instance, the activity coefficients of its salts decrease continuously up to the solubility limit, while the activity coefficients for most common salts decrease at low concentration, but turn around and begin increasing beyond 1 M.[10] Also, it has been shown that $Gnd^+$ is significantly less depleted from the air/water interface than other cations, and it was shown that this cation forms like-charged pairs in aqueous solutions via MD simulation studies.[11,12] [13,14] $Gnd^+$ salts are widely used in biotechnology. Curiously, only guanidinium salts formed with more weakly hydrated counter anions (e.g., GndSCN or GndCl) are used for protein denaturation.[13,15] The mechanism and kinetics of the denaturing action of GndCl on proteins attracted significant recent attention, providing strong experimental[14,16,17] and simulation[17,18] evidence for favorable interaction with most amino acid sidechains.[14,17] There was less consensus, however, as to whether and how strongly this cation interacted with the protein backbone.[19-21] By contrast, salts formed with strongly hydrated anions (e.g. $Gnd_2SO_4$) can actually behave as protein stabilizers.[22] To date, the molecular level mechanism of this counteranion dependency is not well understood.

Herein, we have systematically investigated the influence of guanidinium salts (GndSCN, $GndNO_3$, GndCl, $Gnd_2SO_4$, and $Gnd_2CO_3$) on polypeptide aggregation and hydrophobic collapse in aqueous solutions, focusing on the molecular mechanisms that give rise to protein solubility and stability. ELPs were employed as model biomacromolecules. The polymers consisted of 120 repeating pentameric units of Val-Pro-Gly-Val-Gly with short leader and trailer sequences.[24] This polypeptide displays an inverse phase transition above its lower critical solution temperature (LCST). Previously, it was shown that the LCST behavior of ELPs followed a direct anionic Hofmeister series for a series of sodium salts.[25] Strikingly, however, in the current study we found that a series of $Gnd^+$ salts displayed far richer and more complex phase behavior. Namely, the LCST decreased at concentrations below 0.5 M in the following order: $GndSCN > Gnd_2SO_4 > Gnd_2CO_3 > GndNO_3$. The LCST curve, however, turned around and began to increase for GndSCN above 1 M salt. Moreover, GndCl did not lead to a decrease in the LCST even at low salt concentrations, but continuously caused an increase in the LCST at all concentrations tested.

To understand the origins of macromolecular solubility in the presence of various $Gnd^+$ salts, we performed attenuated total reflection (ATR)-Fourier transform infrared spectroscopy experiments on ELP solutions where the temperature was set above the LCST value. This allowed us to determine whether salt ions were accumulated into or excluded from the collapsed state of the macromolecules. Two diverse salting-out mechanisms for guanidinium salts were discovered that were directly dependent on the identity of the counteranion. Specifically, the polypeptide could be salted out of solution by the standard excluded volume (or 'depletion') mechanism.[23][26] This was the case with $Gnd_2SO_4$ and $Gnd_2CO_3$ salts (Figure 1A). A novel mechanism, however, was discovered based on the accumulation of both $Gnd^+$ and the counteranion into the collapsed state of the polypeptide for GndSCN at low salt concentrations (up to 1.5 M) (Figure 1B). At higher concentrations of GndSCN, the cation and anion coated the polymer surface leading to resolubilization of the macromolecule in an extended conformation. Such intriguing polypeptide solubility data were further rationalized by a combination of a solution theory-based thermodynamic model[27] and atomistic and coarse-grained simulations.[23]

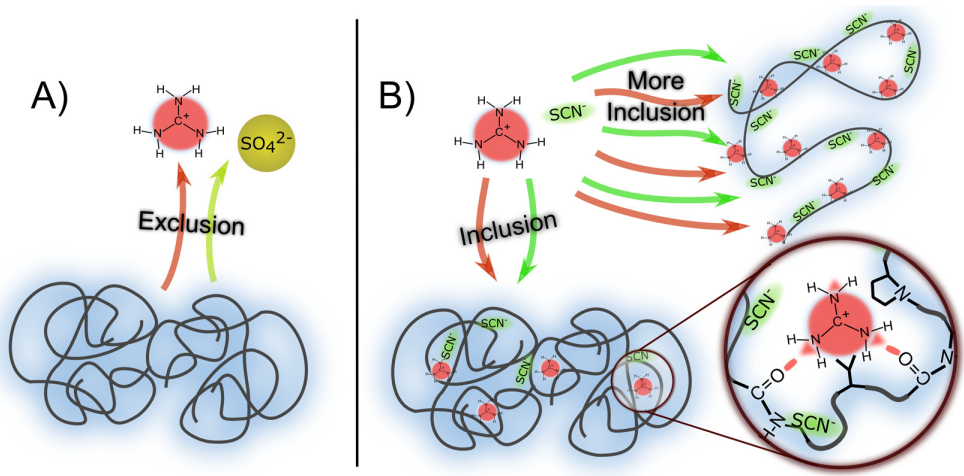

*Figure 1.* (A) Schematic illustration of how Gnd$_2$SO$_4$ makes the collapsed state of the polypeptide favorable via ion exclusion. (B) Schematic illustration of how GndSCN makes the collapsed state of polypeptide more favorable (at <1 M salt concentration) via guanidinium inclusion along with thiocyanate binding to the backbone as illustrated in the zoomed-in picture. This behavior switches over to favoring the uncollapsed polypeptide upon additional ion inclusion at higher salt concentration (> 1.5 M).

## Results

*Phase Transitions of ELPs with Guanidinium Salts.* Figure 2 plots the lower critical solution temperature (LCST) of the ELP with increasing concentrations of 5 Gnd$^+$ salts. As can be seen, substantially different macromolecule solubility trends were observed depending on the salt identity. More specifically, the hydrophobic collapse temperature of the ELP increased linearly as a function of GndCl concentration (green circles), whereas it remained nearly unchanged with GndNO$_3$ (black squares). On the other hand, Gnd$_2$CO$_3$ and Gnd$_2$SO$_4$ decreased the LCST of the polypeptide, where the latter was slightly more efficient in this process (gray diamonds and blue triangles, respectively). All of these salts yielded essentially monotonic trends as a function of salt concentration up to 2M or up to their solubility limit. By contrast, non-monotonic behavior was obtained in the presence of GndSCN (red triangles). Specifically, the polypeptide was salted-out at low salt concentrations, but the solubility increased at higher salt concentrations (>1.5 M). As such, the data suggested that GndSCN had a complex macroscopic effect in which at least 2 major factors influencing the macroscopic behavior were dominant at different concentrations. The overall salting out efficacy at low salt concentrations was as follows:

$$\text{GndSCN} > \text{Gnd}_2\text{SO}_4 > \text{Gnd}_2\text{CO}_3 > \text{GndNO}_3 > \text{GndCl}$$

The place of GndSCN in this series changed and it became a potent salting-in agent at higher salt concentrations (> 1.5 M). This ordering and behavior stand in sharp contrast to that found with both the standard and inversed anionic Hofmeister series.[25] It should be noted that the data in Figure 2 were taken down to 4 °C, and the broken dashed line in the fit to the SCN⁻ curve (shown in red) indicates that no data was obtained between 1.0 and 1.5 M NaSCN.

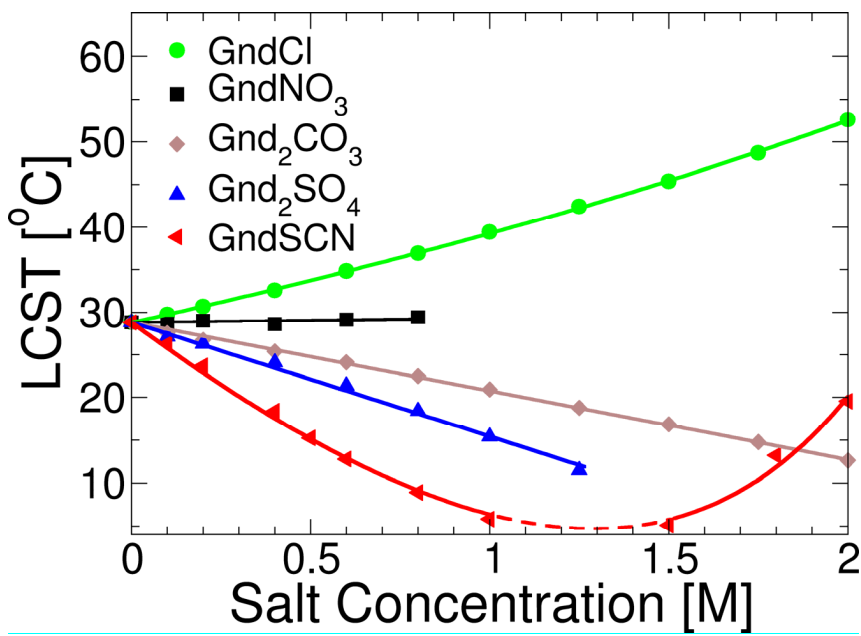

*Figure 2*. LCST measurements of 10 mg/ml ELP solutions as a function of guanidinium salt concentration. All standard deviations were within the data points drawn. Each symbol represents data points from six measurements and the solid lines are fits to eqn. 3. No data were obtained between 1.0 M and 1.5 M GdnSCN, where the LCST value fell below 4 °C (dashed portion of the red line).

Previous LCST measurements with the same ELP, but using sodium salts could be fit to a simple empirical equation which contained just a linear term and a Langmuir isotherm.[7,25,28] This equation, however, could not be used for the GndSCN data in Figure 2. Moreover, a direct Hofmeister series for anions was not observed when Gnd$^+$ was used in place of Na$^+$. This directly raises the question as to whether the standard ideas of binding and depletion are sufficient to understand the phase behavior of the ELPs when Gnd$^+$ is introduced. Specifically, it is generally assumed that ions which salt-out thermoresponsive polymers, like ELPs, would be excluded from the collapsed state. The data in Figure 2 for GndSCN suggest, however, that Gnd$^+$ might induce polymer collapse by accumulating into the collapse state of the macromolecule at low GndSCN concentrations. A recently developed thermodynamic model has correlated protein solubility to a thermodynamic characterization of the differences between the soluble and collapsed states of the macromolecule.[27] Such work predicts a wider variety of more complex LCST line shapes with salt concentration than have previously been observed.

A basic categorization of the possible regimes of salt action with salt type and concentration have been made by statistical mechanics modeling using Flory theory and coarse grain simulations.[23] Three distinct regimes of the concentration dependent cosolvent (e.g., salt) action were observed. The first regime involves collapse due to cosolvent depletion (or exclusion). Under these conditions, the cosolvent is depleted from the polymer/water interface, and the polymer collapses to expose less surface area (a cosolvent entropic effect). In the second regime, the polymer swells due to weak attraction (weak binding). In this regime, the polymer swells to expose a larger surface area because of favorable interactions with the cosolute. Finally, in a third regime, one encounters collapse due to strong attraction (strong binding). In this last case, the cosolute binds strongly enough to the polymer surface that it leads to bridging or crosslinking. Specifically, the cosolvent links to the polymer chains together and thus stabilizes the collapsed state on enthalpic grounds. Guided by these ideas, we decided to directly test the assumption that $Gnd^+$ salts were really always excluded from the collapsed state of ELPs, especially when the introduction of salt led to hydrophobic collapse (e.g. GndSCN at low concentration). This was done using ATR-FTIR. As described in the next section, the vibrational resonances of individual ions were probed and it was possible to determine if they were accumulated into or depleted from the collapsed state of the macromolecules.

**Probing the collapsed state of ELPs by ATR-FTIR.** Several thermoresponsive polymers, including solutions of ELPs, undergo an inverse phase transition above their LCST to form an aqueous two-phase system (ATPS).[29-31] Namely, when the polymer solution is kept above its LCST, the solution separates into protein rich and protein poor phases. For ELPs, the protein rich phase consists primarily of an aqueous solution of the collapsed state of the macromolecule.[29] One can therefore probe whether salt ions are accumulated into this phase or depleted from it. To answer this question, a comparison was made for the absorbance of vibrational bands of the ions in the ELP rich phase with an aqueous salt solution without the ELP. The salt concentration was initially identical in both solutions before phase separation in the ELP solution took place. Details of the measurements and additional controls are provided in the supporting information section (Figure S1).

In a first set of experiments, a 10 mg/ml ELP solution in D$_2$O containing 0.5 M perdeuterated guanidinium thiocyanate (*d*-GndSCN) was heated to 45 °C. The solution phase separated into an ATPS with the protein rich phase forming directly adjacent to a diamond-coated ZnSe crystal that served as a single bounce ATR-FTIR stage. A schematic depiction of this experiment is shown in Figure 3A. Vibrational measurements were made and the resulting spectrum in the amide stretch region is depicted with the green curve in Figure 3B. This spectrum could be fit to a C-N stretching band for the guanidinium cation at 1592 cm$^{-1}$ (red curve) as well as bands from the polypeptide (peaks at 1619 cm$^{-1}$, 1643 cm$^{-1}$ and 1664 cm$^{-1}$). The individual fits for the three polypeptide peaks are shown with gray lines. The first and third peaks are associated with the amide I bands and can be assigned to *β*-turn and *β*-aggregate structures, while the 1644 cm$^{-1}$ band arises from a combination of random coil and distorted *β*-sheet structures. [32][33] Next, the experiment was repeated without any ELP in the solution, but also at 45 °C as a reference spectrum. In this case, only a single peak at 1596 cm$^{-1}$ was found (blue curve). Strikingly, its amplitude was reduced significantly compared to the one with the polymer in the solution. Such a result is strong evidence that the Gnd$^+$ cation is favorably accumulated into the collapsed state in comparison to its uniform concentration in a 0.5 M aqueous solution of *d*-GndSCN. In addition to obtaining data in the C-N stretch and amide band region, data was also taken near 2060 cm$^{-1}$ for the C=N stretch of the SCN$^-$ anion (Figure 3C). As can be seen, SCN$^-$ was also strongly partitioned into the protein rich phase (red curve) compared to its uniform distribution in a similar solution without the ELP (blue curve). It should be noted that both the C-N peak from Gnd$^+$ and the C=N peak from SCN$^-$ were red shifted by ~4 cm$^{-1}$ in the protein rich ATPS phase compared with the resonances found in aqueous solution without the polypeptide. This red shift is direct evidence for strong interactions between the salt ions and the macromolecules.

In a second set of experiments, the partitioning of salt ions into the collapsed state of the ELP was measured in 0.5 M *d*-Gnd$_2$SO$_4$ (Figure 3D). Compared to the results in Figure 3B, the 1643 cm$^{-1}$ peak was reduced in intensity relative to the 1664 cm$^{-1}$ and 1619 cm$^{-1}$ peaks. This is a clear indication of more ordered secondary and tertiary structure for the ELP in this latter case.[34] Moreover, there was far less Gnd$^+$ associated with the protein rich layer compared with the GndSCN experiments. Indeed, an aqueous system containing 0.5 M Gnd$_2$SO$_4$ in the absence of the ELP reveals a somewhat stronger C-N stretch intensity (blue curve) compared to when the biopolymer was present (red curve). As such, Gnd$^+$ was actually modestly depleted from the collapsed state of the polymer. Next, one can look at the S-O stretch resonance at 1098 cm$^{-1}$ from the SO$_4^{2-}$ counteranion (Figure 3E). As can be seen, SO$_4^{2-}$ was strongly depleted from the collapsed polymer state by comparing the S-O resonance intensity (red curve) with the case where the polymer was not present (blue curve). It should be noted that two small bands from the polypeptide near 1046 cm$^{-1}$ and 1060 cm$^{-1}$ were present in the sulfate ion spectrum. These are marked with asterisks to denote that they are from the biopolymer rather than the ion. Significantly, for Gnd$_2$SO$_4$, there was almost no peak shift for either the Gnd$^+$ or SO$_4^{2-}$ resonances in the polymer rich phase compared with aqueous solutions containing no polymer. This would suggest that these ions interacted more weakly with the ELP.

In the final set of vibrational experiments, we tested whether Gnd$^+$ from 0.5 M *d*-GndCl partitioned to the collapsed state of the polypeptide using the amide I spectral region (Figure 3F). Interestingly, the amide I band intensity in the protein rich phase was only about ~45% as high compared to the other two guanidinium salts (Figure 3B&D). This indicates that the density in the protein rich phase was lower, which may reflect a slight charging of the polymer (see SI Section S1 for further details as well as Figure S2E). Moreover, there was a small decrease in the C-N vibrational resonances of Gnd$^+$ ion (1594 cm$^{-1}$) for the solution without the polypeptide. This indicates a very slight partitioning of Gnd$^+$ ions to the collapsed ELPs. There is also a small (~2 cm$^{-1}$) red shift in this vibrational resonance compare to its solution without the ELP, indicating modest GndCl – macromolecule interactions. Unfortunately, the Cl$^-$ partitioning into the collapsed ELPs could not be probed using this method as this anion did not have a corresponding vibrational signature that could be measured. Nevertheless, based on the monotonic increase in the LCST found in Figure 2 as a function of GndCl concentration, modest anion accumulation would be expected to occur. The other two guanidinium salts, GndNO$_3$ and Gnd$_2$CO$_3$, showed analogous trends to the spectra described above (see Figure S2). Namely, Gnd$^+$ ions from GndNO$_3$ accumulated in the protein rich phase, like GndSCN, albeit to a much lesser extent. Moreover, Gnd$_2$CO$_3$ behaved similarly to Gnd$_2$SO$_4$ and the cations remained excluded from the collapsed state of macromolecule (see SI Section S1 and Figure S2).

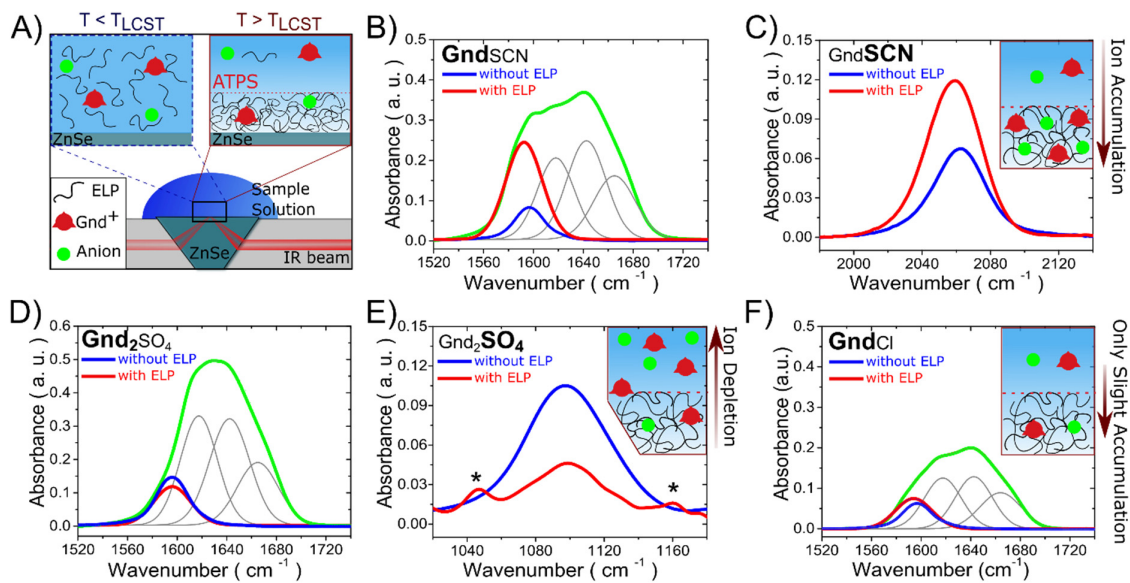

*Figure 3.* A) Experimental scheme showing an aqueous solution droplet (100 μL) placed onto the diamond coated ZnSe ATR crystal. The upper left insert represents an initially homogeneous polypeptide solution (below the LCST), whereas the upper right inset, represents the ATPS which is formed above the LCST. (B) Fitted ATR-FTIR spectra of the collapsed state of the ELP above the LCST (45 °C) in the presence of 0.5 M *d*-GndSCN, in $D_2O$ in the amide I spectral region, and (C) in the C≡N stretch band region of $SCN^-$. (D) Fitted spectra of the same solution except in the presence of 0.5 M *d*-$Gnd_2SO_4$ in $D_2O$ in the amide I spectral region, along with (E) the vibrational spectra of the S-O stretching band of $SO_4^{2-}$. (F) Plots of the amide I spectral region with and without the macromolecule in solutions containing 0.5 M *d*-GndCl. In B-F, the red and blue curves indicate data taken in the presence of the ELP and in its absence, respectively. The gray lines represent three Gaussian fits to the amide I bands, whereas the green curves represent the overall measured spectra. The inset schematics in panels C, E and F depict ion accumulation for GndSCN and ion depletion for $Gnd_2SO_4$ along with only slight ion accumulation for GndCl. The * in (E) denotes weak fingerprint vibrational resonances related to polypeptide.

**Thermodynamic model, coarse-grained and all-atom simulations:** As noted above, classical thermodynamic models would predict that salts which lower the LCST of ELPs would do so via an excluded volume effect. Therefore, one might expect that the ions from 0.5 M solutions of both GndSCN and $Gnd_2SO_4$ would be strongly excluded from the collapsed state of the macromolecules. The data in Figures 2 & 3 reveal that this is not the case for the former salt. Since only $Gdn_2SO_4$ is depleted from the collapsed state, while GndSCN is actually enriched in it, the two salts must lead to hydrophobic collapse by two distinct mechanisms. A new model is therefore required which goes beyond standard mean field descriptions (described in the SI Section S2) and additive models[35-37] to explain the differences between the two salts.

Our thermodynamic model provides a direct link between salt-induced changes in the LCST and three macroscopic thermodynamic parameters of polymer molecules in salt solutions.[27,38] Briefly, this model invokes the observation that the free energy difference of the compact globule ⇌ random coil transition is, at equilibrium, proportional to the logarithm of the population ratio of these two states (equation 1):

$$\Delta G = \mu_{coil} - \mu_{globule} = G_{coil} - G_{globule} = -RT\ln\frac{[Coil]}{[Globule]} \quad (eq.\,1),$$

where R is the universal gas constant and T the absolute temperature. Under a small perturbation, i.e., upon addition of a cosolvent or variation of temperature, the free energy change at the phase transition can be expanded into a Taylor series (equation 2),

$$\Delta G(T_0 + \Delta T, c_s) = \Delta G(T_0, 0) + \left(\frac{\partial \Delta G}{\partial c_s}\right)_T c_s + \left(\frac{\partial \Delta G}{\partial T}\right)_{c_s} \Delta T + \dots \quad (eq.\,2).$$

Significantly, the transition entropy, $\Delta S_0 = -\left(\partial \Delta G / \partial T\right)_{c_s}$, and the m-value, $m = -\left(\partial \Delta G / \partial c_s\right)_T$, fall out naturally as first order responses to these two perturbations (see full derivation in the SI). Since $\Delta G = 0$ at the phase transition temperature, the Taylor series ties the two perturbations together. A particularly useful expression for the change in LCST, $\Delta T(c_s)$, is given by the relation in eqn. 3, which takes the above Taylor expansion out to second order, but sets the heat capacity effects equal to zero, since changes in the LCST are quite small ($\Delta T \ll T_0$).[27] In this case, one gets *m'* and $\Delta S'_0$ terms, leading to the relationship below for the change in the transition temperature:

$$\Delta T(c_s) = -\frac{mc_s + \frac{1}{2}m'c_s^2}{\Delta S_0 + \Delta S'_0 c_s}, (eq.\,3)$$

Results of the application of this model to the LCST data in Figure 2 are presented in Table 1. The two distinct mechanisms for the $SCN^-$ and $SO_4^{2-}$ counteranions are reflected in the values presented in the table. In particular, sulfate has a negative *m* value and a zero *m'* value, consistent with the fact that it salts out under all conditions. The situation is opposite for GndCl, which has a positive *m* value. This is typical for denaturation agents.[27] The situation is more complex for GndSCN. In this case, a very negative *m* value is overcome by a positive *m'* value, which models the turnover from a salting out to a salting in effect. It should be noted that the nonzero value of $\Delta S'_0$ found for GndSCN is essential for a quantitative description of the steep nonlinear rise of the LCST at higher salt concentrations (>1M), and is indicative of attractive GndSCN-ELP interactions. A final point is that eqn. 3 can be rewritten in terms of preferential binding coefficients[27] as discussed in the SI Method section, and the analysis of the LCST data can be directly related to salt partitioning as summarized in SI Section S2 (see Figure S3 and Table S2).

| Salt | $m$ [J/mol/M] | $m'$ [J/mol/M$^2$] | $\Delta S_0'_{fit}$ [J/mol/K/M] |
|---|---|---|---|
| GndCl | 725 | -150 | 13.5 |
| GndNO$_3$ | 55 | 0 | 0 |
| Gnd$_2$CO$_3$ | -590 | 0 | 0 |
| Gnd$_2$SO$_4$ | -1020 | 0 | 0 |
| GndSCN | -2320 | 2170 | 19.7 |

*Table 1:* Summary of the fitted *m*-value, *m'* values, and change in the transition entropy, $\Delta S_0'$ abstracted from fitting the data in Figure 2 by using eqn. 3. The solid lines in the Figure 2 represent fits to eqn. 3. Note that all values are given per molar of the pentameric repeating unit of the ELP. In the fitting procedure, the transition entropy in neat water, $\Delta S_0 = -75$ J/mol/K was used, as determined by the transition enthalpy from DSC experiments.[39]

To help obtain molecular level insight into the thermodynamic model, we employed coarse-grained Langevin dynamics simulations of a chain of beads (representing a flexible polymer) with implicit solvent (see the Supporting Information Methods section for details).[23] Anions and cations were combined into single cosolute beads, which were used to represent salt-polymer interactions. The cosolute-polymer interactions were tuned to model the effects of different salts. In particular, three cases were considered: a repelled cosolvent (resembling Gnd$_2$SO$_4$), a weakly attracted cosolvent (resembling GndCl), and a strongly attracted cosolvent (resembling GndSCN). These three generic types of interactions have been described in previous work.[23] The results from coarse-grained simulations for each of these cases are presented in Figure 4A. The graph plots the radius of gyration of the polymer as a function of cosolvent concentration, where smaller polymer sizes (radius of gyration values, $R_g$) represent the collapsed state of an ELP, while larger more extended polymers would be equivalent to the uncollapsed state. As expected, the strongly excluded cosolvent simply led to the collapse of the polymer (blue curve), while the weakly attracted cosolvent led to polymer swelling (green curve). Significantly, a strongly attracted cosolvent led to salting-out at low concentration, but salting-in at higher concentration (red curve). These simulations capture the key features found in Figure 2. As expected, an excluded cosolvent led to the collapse of the polymer chain (analogous to Gnd$_2$SO$_4$), while a weakly binding cosolvent (analogous to GndCl) led to polymer swelling. However, the case with a strongly binding cosolvent was more complex (analogous to GndSCN). In this case, collapse was seen at low cosolvent concentrations, but re-entrant swelling was found at higher concentrations.

To understand the origins of these three regimes in more detail, we have analyzed the cosolvent distribution near the polymer center (Figure 4B). Specifically, we have introduced a radial distribution function $g(r) = \rho(r)/\rho_0$. This is the ratio of the local cosolvent density at a distance $r$ from the polymer center, $\rho(r)$, with the average cosolvent density denoted as $\rho_0$. At lower cosolvent concentrations (~1 M), we see striking differences depending on the type of cosolvent which was employed. Namely, the excluded cosolvent (blue curve) was depleted from the vicinity of the vicinity of the polymer ($g(r) < 1$), the weakly interacting cosolvent caused little changed (green curve), while the strongly interacting cosolvent (red curve) led to significant enrichment of the cosolvent ($g(r) \gg 1$). These simulation results strongly mirror the trend found in the ATR-FTIR measurements (Figure 3). On the other hand, at a much higher cosolvent concentrations (~13 M, inset in Figure 4B), the strongly binding cosolvent shows much less pronounced accumulation. This is expected, as even a strongly binding cosolvent will saturate and the $g(r)$ values decrease towards 1, due to the normalization with respect to the bulk cosolvent concentration.

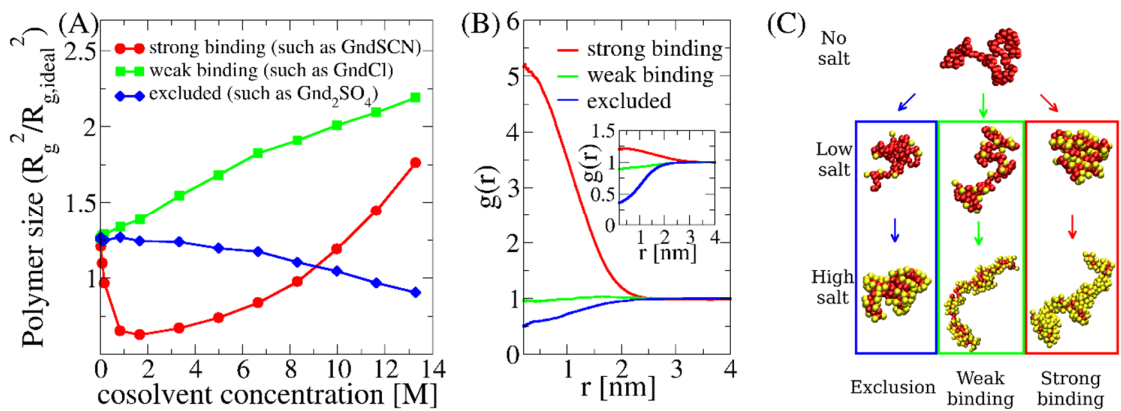

*Figure 4.* (A) Results of coarse-grained simulations for a model polymer in cosolvent solutions that bind strongly (red), weakly (green), or are depleted from the polymer surface (blue).[33] The results plot the radius of gyration of the polymer (scaled by that of an ideal chain with the same bond length) as a function of cosolvent concentration. (B) The distribution of cosolvent from the center of the polymer at 1 M concentration. The inset presents the same distribution curves at 13 M cosolvent. In panel (B), $r$ denotes the distance to center of mass of the polymer. (C) A schematic depiction of the mechanism of swelling and compression of the polymer (red spheres) caused by the cosolvent moieties (yellow spheres) in the low (1 M) and high (13M) concentration regimes. Note that the existence of multivalent binding interactions of the cosolvent at low solvent concentration is responsible for the polymer collapse in the strong binding regime. Only the cosolvent molecules in direct contact (< 4 A) with the polymer chain are depicted for clarity.

Snapshots from the Langevin dynamics simulations of the polymer as a function of cosolvent concentration are shown for the three different binding regimes (Figure 4C). These cosolvent distributions correlated well with the ion distributions of the Gnd$^+$ salts found around the macromolecules using explicit water in all atom MD simulations (described below in Figure 5) as well as the ion related vibrational signals observed in the ATR-FTIR spectra (Figure S2 and Figure 3, respectively). In particular, a change in the conformation of the polymer from a more globular to a more extended structure occurred as the cosolvent concentration was increased in the strong binding regime. This result was consistent with the idea that strongly interacting cosolvents bound the polymer chains together at low concentration. However, as the cosolvent concentration was increased, the polymer chains became coated with the cosolvent and were, thus, extended by it.

Finally, salt-specific affinity to the ELP surface was investigated by all atom MD simulations (see SI Methods for full simulation details). Figure 5 shows Gnd$^+$ distributions (in purple) near a pentameric elastin motif (VPGVP with capping groups on each end) with (A) SO$_4^{2-}$ (in silver), (B) Cl$^-$ (in orange), and (C) SCN$^-$ (in yellow) counterions, respectively. As can be seen, guanidinium cations could interact with the carbonyl groups at the backbone via H-bonding with its amine groups. The hydrophobic side chains were found to interact with the flat CN$_3$ faces of the cations. The differences between the investigated salts are clearly manifested. Gnd$_2$SO$_4$ was found to be the most depleted ion at the peptide surface, while GndCl was neither strongly attracted nor depleted. On the other hand, GndSCN was significantly enriched at the peptide surface. The locations of Gnd$^+$ at the peptide surface were similar for all three salts, but the amount increased in the order: SO$_4^{2-}$ < Cl$^-$ < SCN$^-$.

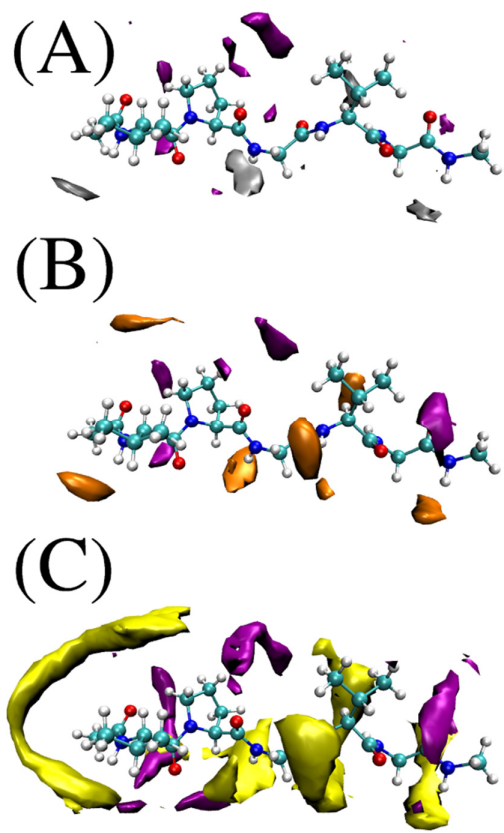

*Figure 5.* Spatial density maps of the ions, as obtained from all atom MD simulations. (A) This snapshot shows the depletion of $Gnd_2SO_4$ ($Gnd^+$ in purple and $SO_4^{2-}$ in silver) from the vicinity of the VPGVG pentapeptide. (B) GndCl ($Cl^-$ in orange) and (C) GndSCN ($SCN^-$ in yellow). The contours plotted for each ion correspond to 4x the bulk density for all three snapshots.

The radial distribution functions corresponding to the snapshots in Figure 5 are provided in Figure 6. The preferential binding coefficients to the extended state of elastin are presented in Figure S8 in the SI. The affinities of the guanidinium salts to elastin differ substantially from each other and can be ordered as $GndSCN > GndCl > Gnd_2SO_4$.

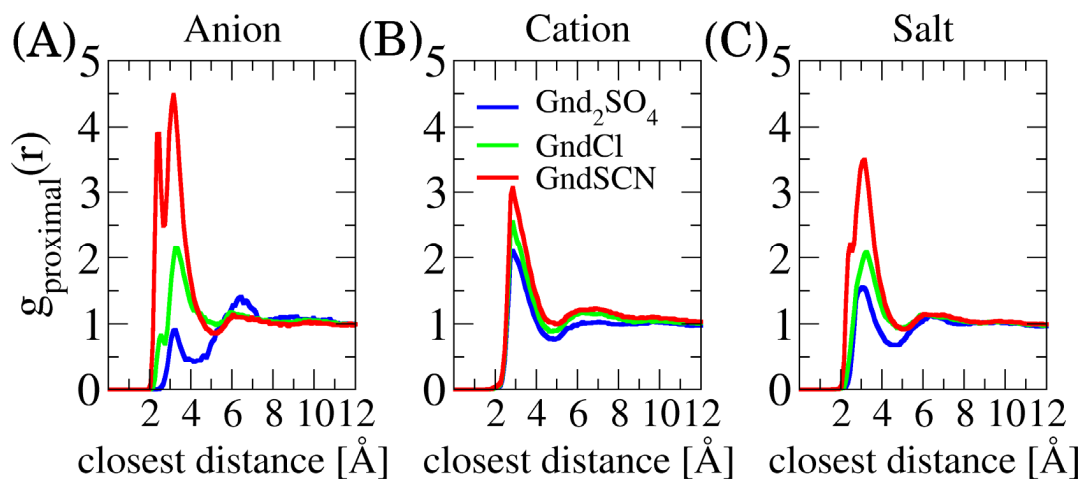

*Figure 6.* Proximal distribution function of the investigated guanidinium salts around the extended VPGVG pentapeptide. The distribution of anions is shown in the panel (A), that of the $Gnd^+$ cation is in the panel (B), and overall salt distribution in the panel (C).

**Discussion and Conclusions**

Gnd$^+$ is a powerful denaturant that interacts with polypeptides in a unique and complex fashion compared to metal cations. Even though metal ions can form ion pairs with aspartate and glutamate residues, they are strongly depleted from the more hydrophobic portions of proteins. Indeed, even alkali earth ions such as Mg$^{2+}$ and Ca$^{2+}$ associate only very weakly with the amide oxygen on the backbones of proteins.[9,40] Gnd$^+$ is different in that it can form multiple donating hydrogen bonds along its edges. Moreover, the hydrophobic faces of this cation can interact with the hydrophobic portions of peptides and proteins. Even so, its interactions with uncharged polypeptides are weak and dictated by the nature of the counteranion in solution with which it is paired. Specifically, when paired with a strongly excluded anion, such as SO$_4^{2-}$, guanidinium is somewhat depleted from the polymer/water interface. This leads to the stabilization of the collapsed state of ELPs via a standard excluded volume effect (Figure 3D&E, and Figure S2). When guanidinium is paired with a more weakly hydrated anion such as SCN$^-$, the situation is changed. SCN$^-$ partitions favorably to the amide backbone of polypeptides and Gnd$^+$ follows the anion and becomes enriched at the polypeptide/water interface (Figure 3B&C). Under these conditions, guanidinium helps to keep the polypeptide in the collapsed state at low concentration as it can bridge polymer chains together via intra- and inter-molecular H-bonding interactions. The collapsed state in this case is quite different than the one caused by salt ion depletion. Indeed, ATR-FTIR reveals that the amide I region in the collapsed peptide chains contain less secondary structure when GndSCN is incorporated compared to conditions where salting out occurs via exclusion. Moreover, at higher GndSCN concentrations, the effect of the salt changes entirely. In that case, a sufficient amount of guanidinium is bound such that individual cations can no longer bridge multiple portions of the polymer chains together. As such, the salt switches over to stabilizing the uncollapsed state of the macromolecule and works as a more classic denaturant. It should be noted, that collapse and re-entrant behavior was also found upon introducing GndSCN to additional thermal response polymer systems (see Section S5 and Figure S9). Such results speak to the generality of this mechanism.

When used as a denaturant, Gnd$^+$ is typically either paired with Cl$^-$ or SCN$^-$.[41,42] The current studies show it would simply be ineffective as a denaturant when paired with a strongly excluded anion such as SO$_4^{2-}$. For proteins which are well above their isoelectric point, Cl$^-$ is a satisfactory choice, as Gnd$^+$ will partition to the negatively charged interface on electrostatic grounds. In this case, it should still cause a stabilizing effect at low salt concentration, but lead to denaturation a high salt concentration. By contrast, proteins that are near to or below their isoelectric points require the use of GndSCN to cause denaturation as $Gnd^+$ is not partitioned favorably to neutral or positively charged polymer surfaces. As such, the partitioning of the anion to the interface is needed to render the protein surface sufficiently negatively charged to recruit $Gnd^+$. It should be noted that NaSCN is already a reasonably good denaturant and $SCN^-$ will certainly help cause denaturation on its own. However, $Na^+$ is partitioned away from the protein and replacing it with $Gnd^+$ creates a far more effective denaturant since both the cation and anion concentration are enhanced at the interface. Nevertheless, an additive partitioning model[36,37] is not really applicable here, since the affinity of GndSCN for peptides is larger than that which would be inferred from the combined affinities of NaSCN and GndCl. Finally, $Gnd^+$ is a more effective denaturant than $NH_4^+$, which interacts only via hydrogen bonding, as it does not partition favorably to the hydrophobic regions of the polypeptide. Also, $NH_4^+$ does not have the curious property of stabilizing hydrophobic collapse at lower salt concentrations, as it is not an effective cross-linker.

**Notes:**

[†]The current address for H.I.O is the Laboratory for Fundamental Biophotonics (LBP), École Polytechnique Fédérale de Lausanne (EPFL), Switzerland.

**Acknowledgement:**

JHe and JD thank the Alexander-von-Humboldt (AvH) Stiftung, Germany, for financial support. PJ thanks the Czech Science Foundation (Grant P208/12/G016) for support. JHe thanks the Czech Science Foundation (Grant 16-57654Y) for support. PSC thanks the National Science Foundation (CHE-1413307).

## Supplementary:

**The supplementary information includes the** Materials and Methods Section, and the sections for ATR-FTIR measurements of ELPs above its LCST, thermodynamic model of the LCST data, coarse grained simulations and applications of the thermodynamic model, MD simulations of ion-macromolecule interactions, and LCST values for other thermoresponsive polymers in GndSCN.

**TOC Figure**

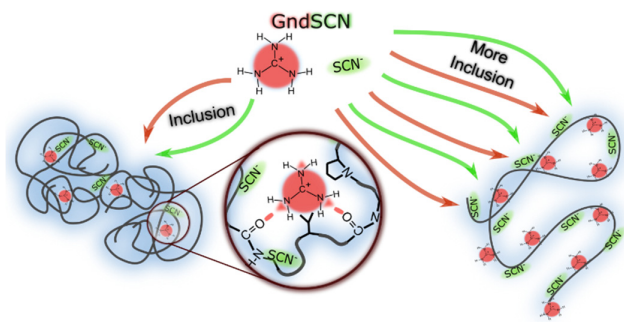